\documentclass[a4paper]{mem}
\usepackage{natbib}
\usepackage{graphicx}
\usepackage{amsmath}
\usepackage[a4paper]{hyperref}
\idline{00}{0}
\catcode`\@=11
\def\gsim{\ifmmode{\mathrel{\mathpalette\@versim>}}
    \else{$\mathrel{\mathpalette\@versim>}$}\fi}
\def\lsim{\ifmmode{\mathrel{\mathpalette\@versim<}}
    \else{$\mathrel{\mathpalette\@versim<}$}\fi}
\def\@versim#1#2{\lower 2.9truept \vbox{\baselineskip 0pt \lineskip
    0.5truept \ialign{$\m@th#1\hfil##\hfil$\crcr#2\crcr\sim\crcr}}}
\catcode`\@=12
\def\msol{M_{\odot}}
\def\lsol{L_{\odot}}
\def\mpc{{\rm Mpc}}
\def\kpc{{\rm kpc}}
\def\rvir{r_{\rm vir}}
\def\rh{R_{\rm h}}
\begin{document}
   \title{High-resolution re-simulations of massive DM halos and the
   Fundamental Plane of galaxy clusters 
%\thanks{this is a place for a title footnote}
}

   \author{B. Lanzoni \inst{1}, 
   A. Cappi \inst{1}, \and L. Ciotti \inst{2}
   }

   \offprints{B. Lanzoni} \mail{Osservatorio Astronomico di
   Bologna,via Ranzani 1, 40127 Bologna } 

   \institute{INAF -- Osservatorio Astronomico di Bologna, via Ranzani
              1, 40127 Bologna, Italy \email{lanzoni@bo.astro.it,
              cappi@bo.astro.it}\\   
    \and Dipartimento di Astronomia, Universit\`a di Bologna, via
              Ranzani 1, 40127 Bologna, Italy
              \email{ciotti@bo.astro.it} \\ 
                           }

\abstract{Pure N-body high-resolution re-simulations of 13 massive
dark matter halos in a $\Lambda$CDM cosmology are presented. The
resulting sample is used to investigate the physical origin of the
Fundamental Plane, ``Faber-Jackson'', and ``Kormendy'' relations
observed for nearby galaxy clusters. In particular, we focus on the
role of dissipationless hierarchical merging in establishing or
modifying these relations. Contrarily to what found in the case of the
Faber-Jackson and Kormendy relations for {\it galaxies} (see
Londrillo, Nipoti \& Ciotti, this conference), dissipationless merging
on {\it cluster scales} produces scaling relations remarkably similar
to the observed ones. This suggests that gas dissipation plays a minor
role in the formation of galaxy clusters, and the hierarchical merger
scenario is not at odd with the observed regularity of these systems.

   }
   \authorrunning{B. Lanzoni et al.}
   \titlerunning{High-resolution re-simulations and the cluster FP}
   \maketitle
%
%________________________________________________________________

\section{Introduction}
Like early-type galaxies, also galaxy clusters follow well defined 
scaling relations among their main observables: in particular, 
a luminosity--velocity dispersion relation (i.e., a
``Faber-Jackson--like'' relation), a luminosity--radius relation (i.e., a
``Kormendy--like'' relation), and the  Fundamental Plane (hereafter
FP; e.g., Schaeffer et al. 1993, hereafter S93; Adami et al. 1998; and
for ellipticals: Djorgovski et al. 1987; Dressler et al. 1987; Faber
\& Jackson 1976; Kormendy 1977). 
Besides their potential importance as distance indicators, these 
relations contain information on the cluster formation and evolution
processes.
Within the commonly accepted cosmological scenario, where cold 
dark matter (CDM) is the dominant component of the Universe, structure 
formation is driven by hierarchical dissipationless merging: small DM
halos form first, then they merge together to form larger systems, and
the evolution of the baryonic matter follows that of the hosting DM
halos. 

The importance of investigating the physical origin of the observed
scaling relations, and the role of dissipationless hierarchical
merging in establishing or modifying them, is therefore apparent, and
explains the large number of theoretical works devoted to this problem
in the case of elliptical galaxies (e.g. Capelato et al. 1995; Nipoti,
Londrillo, Ciotti 2002a,b; Gonzales \& van Albada 2002; Evstigneeva et
al. 2002; Dantas et al. 2002; Londrillo et al., this conference).  In
particular, it has been shown that, while the FP is reproduced by the
end--products of equal mass mergers, the Faber--Jackson and the
Kormendy relations are {\em not} (Nipoti et al. 2002ab; Londrillo et
al., this conference).  In the case of galaxy clusters however, almost
no theoretical studies are devoted to their FP (see, e.g., Pentericci,
Ciotti \& Renzini 1995; Fujita \& Takahara 1999; Beisbart, Valdarnini
\& Buchert 2001).

We address this problem in the present work, by means of high
resolution re-simulations of very massive DM halos (the hosts of
galaxy clusters) in a $\Lambda$CDM cosmology.  Following their
hierarchical dissipationless evolution in detail, we verify whether
they reproduce the observed scaling relations, both at the present
time, and at higher redshift.  A full analysis of this work will be
given elsewhere (Lanzoni et al., in preparation), while we present
here a selection of the main astrophysical results, focusing on the
numerical issues.

\section{High-resolution re-simulations}

To investigate whether the dark matter counterpart of galaxy clusters,
as obtained by numerical simulations, do define the observed scaling
relations, a large enough sample of very massive DM halos was
needed. For this purpose, we have employed the dissipationless ``Very
Large Simulations'' (VLS; see Yoshida, Sheth, \& Diaferio 2001), where
the simulated comoving volume of the Universe is sufficiently large:
the box side is of $479\,h^{-1}\mpc$, with $H_0 = 100\,h^{-1}$ km
s$^{-1}$ Mpc$^{-1}$, and $h=0.7$.  The adopted cosmological model is a
$\Lambda$CDM Universe with $\Omega_{\rm m}=0.3$, $\Omega_\Lambda=0.7$,
spectral shape $\Gamma = 0.21$, and normalization to the cluster local
abundance, $\sigma_8=0.9$.  The total number of particles is $512^3$,
of $6.86\times 10^{10}\msol/h$ mass each.

From these simulations, we have selected a sample of 13 halos with
masses between $10^{14}\msol/h$ and $2.3\times 10^{15}\msol/h$.  They
span a variety of shapes, from nearly round to more elongated.  The
richness of their environment also changes from case to case, with the
less isolated halos usually surrounded by pronounced filamentary
structures, containing massive neighbors (up to $20\%$ of their mass).
All the selected halos are required to have a massive progenitor at
redshift $z=0.5$ (their masses range between $6\times 10^{13}$ and
$1.5\times 10^{15}\msol/h$), and also at $z=1$ (with masses between
$4\times 10^{13}$ and $7\times 10^{14}\msol/h$).  Such a choice allows
us to study galaxy clusters not only locally, but also at higher
redshift.

Given the mass resolution of the simulations, less than 1500 particles
compose a halo of $10^{14}\msol/h$, and its properties entering the FP
relation cannot be accurately determined.  We have therefore
resimulated at higher resolution the halos in our sample, by means of
the technique introduced by Tormen, Bouchet \& White (1997), that we
briefly summarize in the following.

The first step is to select, in a given cosmological simulation, the
halo one wants to reproduce at higher resolution.  Then, all the
particles composing the selected halo and its immediate surroundings
are detected in the initial conditions of the parent simulation: the
number of particles within the region thus defined is increased, and
the region is therefore called ``the high resolution (HR) region''.
As a consequence, the mean inter-particle separation decreases, and
the corresponding high-frequency modes of the primordial fluctuation
spectrum are added to those on larger scales originally used in the
parent simulation. The overall displacement field is also modified
consequently.  At the same time, the distribution of surrounding
particles is smoothed by means of a spherical grid, whose spacing
increases with the distance from the center: in such a way, the
original surrounding particles are replaced with a smaller number of
\emph{macroparticles}, whose mass grows with the distance from the HR
region.  Thank to this method, even if the number of particles in the
HR region is increased, the total number of particles to be evolved in
the simulation remains small enough to require reasonable
computational costs, while the tidal field that the overall particle
distribution exerts on the HR region remains very close to the
original one.  For the new initial configuration thus produced, vacuum
boundary conditions are adopted, i.e., we assume a vanishing density
fluctuation field outside the spherical distribution of particles with
diameter equal to the original box size $L$.  A new N-body simulation
is then run starting from these new initial conditions, and allows to
re-obtain the selected halo at the suited resolution.

We have applied this technique to the 13 massive DM halos selected in
the VLS.  For 8 out of the 13 selected halos, the resolution has been
increased by a factor $\sim33$, using high-resolution particles of
$\sim2.07\times 10^9\msol/h$ each, while a factor of 2 improvement has
been adopted for the 5 intermediate mass halos (the particle mass is
$10^9\msol/h$ in this case).  The gravitational softening used for the
high-resolution region is $\epsilon = 5\,\kpc/h$, corresponding to
about $0.2\%$ and $0.5\%$ of the virial radius of most and less
massive halos, respectively. This scale length represents the spatial
resolution of the resimulations, to be compared with that of
$30\,\kpc/h$ of the original VLS.  Note that to prevent low-resolution
macroparticles to ``contaminate'' the final resimulated halo (i.e., to
end within its virial radius), we need to define the HR region not
only through the particles of the halo itself in the parent
simulation, but considering also a boundary region in its immediate
surroundings.  This is particularly important in the case of less
massive halos, since the small number of their composing particles
does not allow to define their encompassing region in the initial
conditions with enough precision.  To avoid such a ``contamination''
problem in our re-simulations, a radius of about $3\,\rvir$ and
$5\,\rvir$ for the boundary region was necessary for the most and the
less massive halos, respectively.  The resulting number of high and
low resolution particles are listed in Table \ref{tab:simu}, together
with their sum\footnote{Note that to get the same mass resolution
($2.07\times 10^9\msol/h$ or $10^9 \msol/h$) in the entire volume of
the parent simulation, about $1650^3$ of $2100^3$ particles of would
have been required.}.  To run the resimulations, the parallel
dissipationless tree-code ``GADGET'' (Springel, Yoshida \& White 2001)
has been used on the IBM SP2 and SP3 of the Centre Informatique
National de l'Enseignement Sup\'erieur (CINES, Montpellier, France),
and the CRAY T3E of the RZG Computing Center (Munich, Germany), that
have comparable processor speed.  The number of processors and the CPU
time required for the corresponding runs are also listed in Table
\ref{tab:simu}.

\begin{figure*} \centering
   \resizebox{\hsize}{!}{\rotatebox[]{0}{\includegraphics{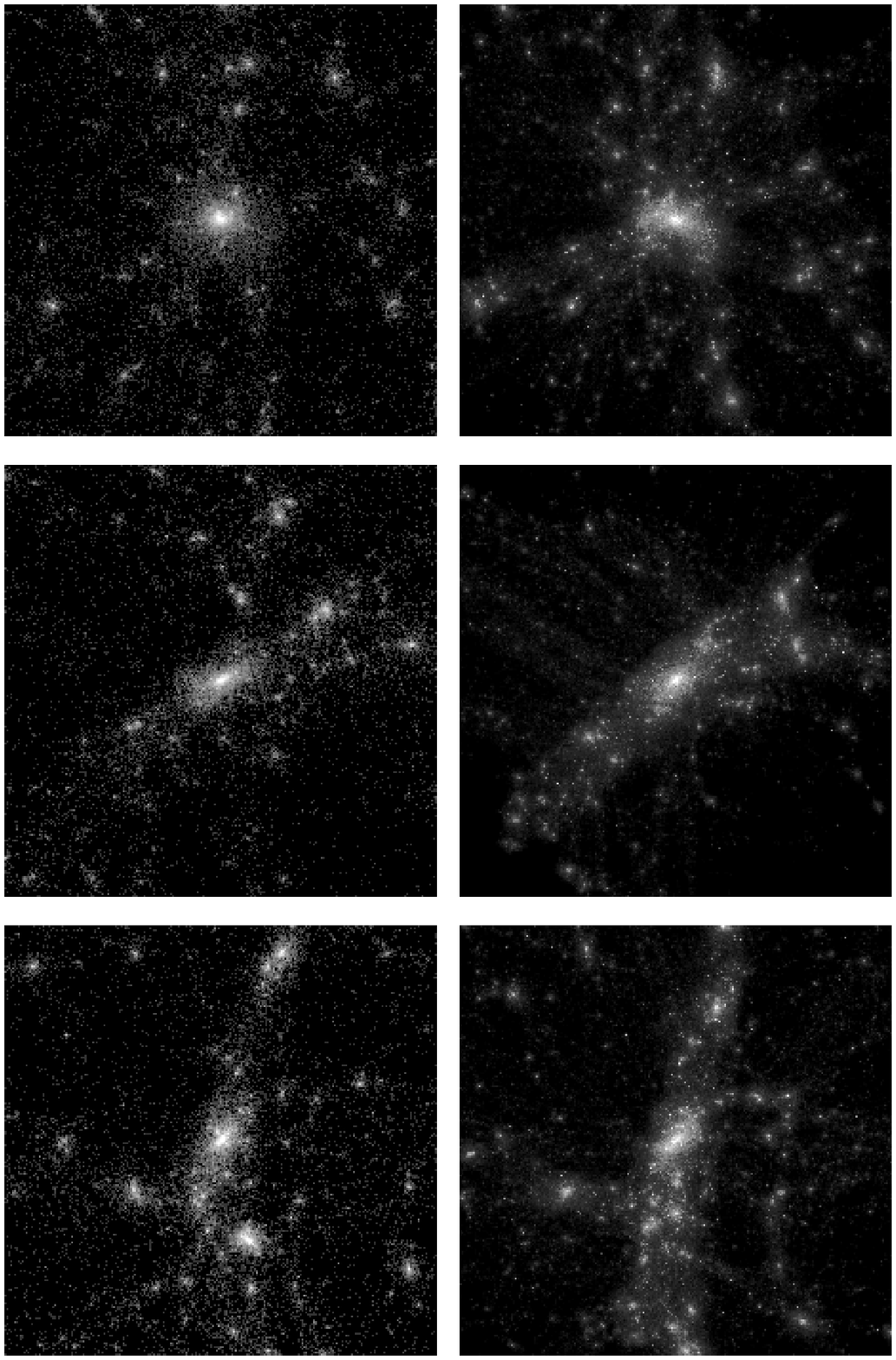}}}
   \caption{Images of the 3 most massive DM halos in the parent
   simulation (left panels) and after the high-resolution
   re-simulation (right panels). All panels show the projection along
   an arbitrary line of sight, of a region of $\sim 30\,\mpc/h$ around
   the halo center of mass.}  
   \label{fig:ima30mpc}
\end{figure*}

\begin{figure*} \centering
   \resizebox{\hsize}{!}{\rotatebox[]{0}{\includegraphics{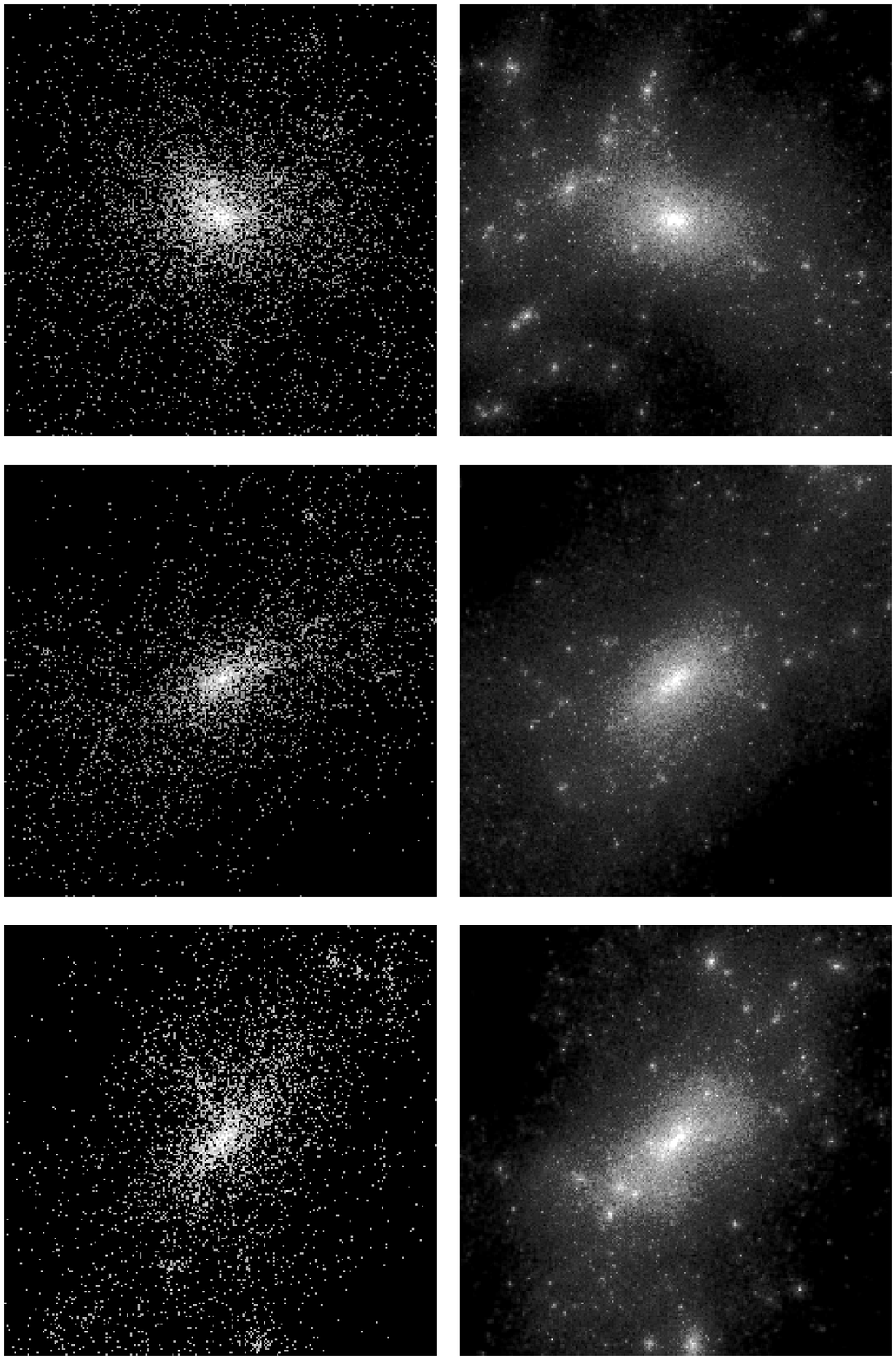}}}
   \caption{Images of the 3 most massive DM halos in the parent
   simulation (left panels) and after the high-resolution
   re-simulation (right panels). All panels show the projection along
   an arbitrary line of sight, of a region of $\sim 2\,\rvir$ around
   the halo center of mass.}
   \label{fig:ima2rvir}
\end{figure*}

Results have been dumped in output files for 100 time steps, equally
spaced in the logarithm of the expansion factor of the Universe
between $z=20$ and $z=0$.  At each time step, a Spherical Overdensity
(Lacey \& Cole 1994) halo finder has been used to detect the halos,
and to estimate their virial mass and radius (see Table
\ref{tab:simu}).  With respect to the original virial masses and
radii, those of the resimulated halos show non-systematic and small
differences (of the order of few percent), particularly for the most
massive objects.  Moreover, also the halo formation history (the way
how mass assembly proceeds with time) is very similar to that in the
parent simulation, thus assuring the reliability and precision of the
high-resolution re-simulation technique we use.  Visually, the
noticeable improvement provided by such a technique is apparent from
Figs. \ref{fig:ima30mpc} and \ref{fig:ima2rvir}, where the images of
the 3 most massive halos, before and after the resimulations, are
shown at the scales of $\sim 30\,\mpc/h$ and $\sim 2\,\rvir$,
respectively.

%-------------------------------------------------------------_
\begin{table*}
\caption[]{Characteristics of the high-resolution resimulations of
the 13 massive halos: N$_{\rm HR}=$ number of high-resolution
particles, N$_{\rm LR}=$ number of low-resolution macroparticles,
N$_{\rm T}=$ total number of particles, N$_{\rm proc}=$ number of
processors, $t_{\rm CPU}=$ hours per processor required for the
resimulations, M$_{\rm vir}=$ virial mass in $10^{14}\msol/h$, $r_{\rm
vir}=$ virial radius in Mpc/$h$.} 
\label{tab:simu} 
\halign{\strut# & #\hfil &\quad\hfil
                  #\hfil &\quad\hfil
                  #\hfil &\quad\hfil
                  #\hfil &\quad\hfil
                  #\hfil &\quad\hfil
                  #\hfil &\quad\hfil
                  #\cr 
 \noalign{\hrule \vskip 2 pt} \cr
 Name  & N$_{\rm HR}$  & N$_{\rm LR}$ & N$_{\rm T}$ & N$_{\rm proc}$ &
$t_{\rm CPU}$ & M$_{\rm vir}$ & $r_{\rm vir}$\cr 
\noalign{\vskip 5 pt \hrule \vskip 7 pt} 
g8    & 3700120 & 191733 & 3891853 & 32    & 80   & 23.42 & 2.75 \cr
g1    & 2574717 & 202301 & 2777018 & 32    & 70   & 13.99 & 2.31 \cr
g72   & 3299865 & 194277 & 3494142 & 32    & 57   & 11.77 & 2.18 \cr
g696  & 4870197 & 184314 & 5054511 & 64    & 60   & 11.37 & 2.16 \cr
g51   & 1677364 & 213477 & 1890841 & 32    & 44   & 10.78 & 2.12 \cr
g245  & 3437317 & 215269 & 3652586 & 16    & 18.4 &  6.50 & 1.79 \cr
g689  & 3252085 & 215809 & 3467894 & 16    & 18.3 &  6.08 & 1.75 \cr
g564  & 2068981 & 227187 & 2296168 & 16    & 12.8 &  4.91 & 1.63 \cr
g1777 & 3094123 & 219047 & 3313170 & 16    & 18.0 &  3.83 & 1.50 \cr
g4478 & 2293433 & 225706 & 2519139 & 16    & 12.8 &  2.92 & 1.37 \cr
g914  &  250605 & 247091 &  497696 & 16    & 7.3  &  1.45 & 1.09 \cr
g3344 &  206140 & 248756 &  454896 & 16    & 6.3  &  1.09 & 0.99 \cr
g1542 &  207202 & 248948 &  456150 & 16    & 6.3  &  0.82 & 0.99 \cr
\noalign{\vskip 5 pt \hrule \vskip 2 pt}
\cr}
\label{tab:kvir}
\end{table*}
%-------------------------------------------------------------

\section{Observed scaling relations at $z=0$}

Analyzing a sample of 16 nearby galaxy clusters, S93 found the
following relation between the observed luminosity and velocity
dispersion:
\begin{equation}
L\propto \sigma^{1.87\pm 0.44},
\label{eq:fj}
\end{equation}
where $L$ is in units of $L_* = 1.325 \times 10^{10}\,h^{-2} \lsol$
($h=1$), and $\sigma$ in km/s, and between luminosity and radius:
\begin{equation}
L\propto R^{1.34\pm 0.17}, 
\label{eq:lr}
\end{equation}
where $R$ is the effective radius measured in Mpc.  In both cases, the
observed scatter is very large, while a considerable improvement is
found when linking the three observables together in a FP-like
relation.  While a bi-parametric fit was performed by S93, here we
derived the FP by means of the Principal Component Analysis (hereafter
PCA; see e.g. Murtagh \& Heck 1987), that searches for the most
suitable combinations of the three observables able to describe a thin
plane.  By performing a PCA on the data sample of S93, we obtain the
new orthogonal variables
\begin{equation}
p_i\equiv\alpha_i \log R + \beta_i\log L + \gamma_i \log\sigma, 
\label{eq:pi}
\end{equation}
with $R$ in Mpc/$h$, $L$ in $100L_*$, $\sigma$ in 1000 km/s.  In
Fig. \ref{fig:fp} we show the resulting distribution of the observed
clusters (blue triangles) in the ($p_1,p_3$) and ($p_1, p_2$) spaces,
the former providing an exact edge-on view of the FP and making
apparent its small thickness.

%------------------------------------------Two column figure (place early!)
%______________________________________________________________

%------------------------------------------Two column figure (place early!)
%______________________________________________________________

\section{Scaling relations at $z=0$ for simulated clusters}

To derive the analogous quantities for the simulated clusters, we have
constructed their projected radial profiles by counting the DM
particles within concentric shells around the center of mass, for
three orthogonal directions.  Thus, the projected half-mass radius
$\rh$ has been derived as the radius of the shell containing half the
total number of particles, and the velocity dispersion $\sigma$ has
been computed by averaging the squared line of sight component of the
(barycentric) velocity over all the particles within $\rh$.  Since the
simulated clusters, as well as the real ones, are not spherical, such
a procedure gives different values of $\rh$ and $\sigma$ for the three
considered line of sights, the maximum difference never exceeding 33\%
and 21\% in the two cases, respectively.  Finally, for converting the
DM mass into light, we have adopted a mass--to--light ratio with a
small dependence on the luminosity, as suggested by observations (S93;
Girardi et al. 2002): $M/L\propto L^\alpha$, with $\alpha=0.3$. In
particular, we have assumed: $M/L = 450 (M/M_0)^{\alpha/(\alpha+1)}
\msol/\lsol$, where $M_0=7.5\times 10^{14}\msol$.

\begin{figure*} \centering
  \resizebox{\hsize}{!}{\rotatebox[]{0}{\includegraphics{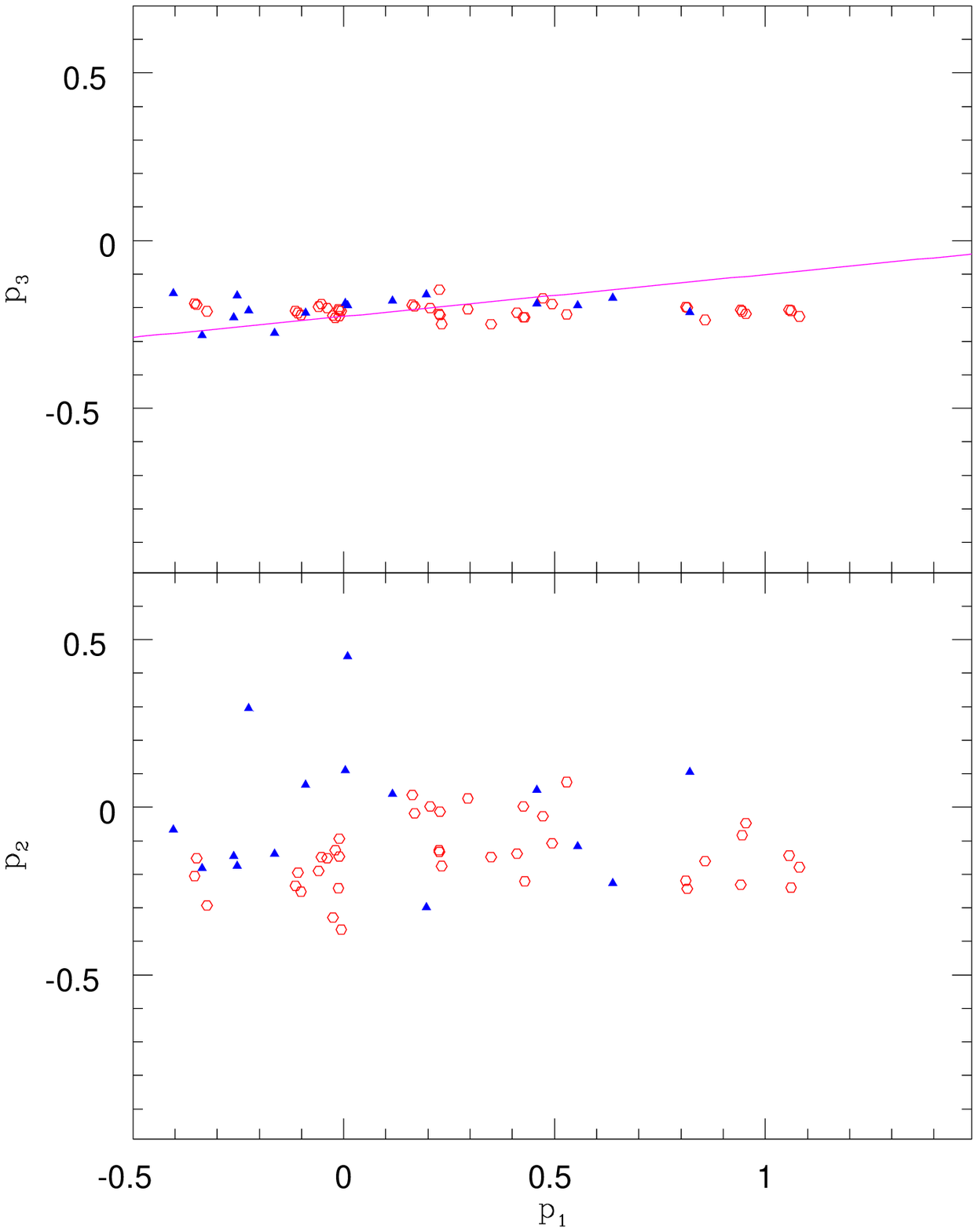}}}
   \caption{Observed (blue triangles) and simulated (red hexagons) 
   clusters in the edge-on and face-on view of the FP (upper and
   lower panels, respectively), at redshift $z=0$. The linear best fit
   to the simulated clusters in case of a constant $M/L$ is shown by
   the pink solid line.}
   \label{fig:fp}
\end{figure*}

\begin{figure*} \centering
  \resizebox{\hsize}{!}{\rotatebox[]{0}{\includegraphics{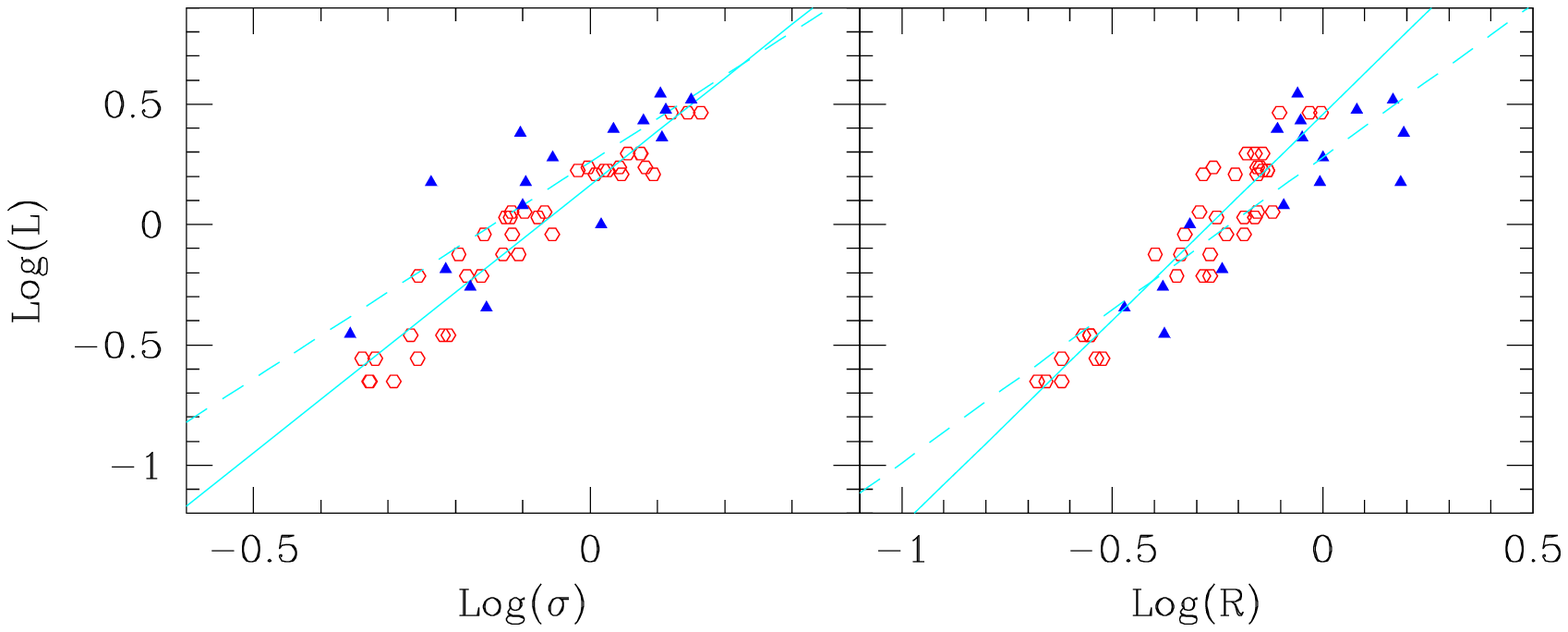}}}
   \caption{$L$-$\sigma$ (left panels) and $L$-$R$ (right panels)
   relations for the observed (blue triangles and dashed line) and
   simulated (red hexagons and solid line) clusters, at redshift
   $z=0$. Luminosities are in units of $100\,L_*$, velocity
   dispersions in km/s, and radii in Mpc.}
   \label{fig:fj}
\end{figure*}

By performing a PCA on the 13 DM halos for the three considered lines
of sights, using $\rh$, $\sigma$, and $L$ computed as just described,
we obtain a very well defined and thin FP.  To check whether this FP
is similar to the observed one, we have also constructed the PCA
variables $p_i$ by combining $\rh$, $\sigma$, and $L$ derived for the
DM halos, with the coefficients $\alpha_i$, $\beta_i$, $\gamma_i$
obtained for the observed clusters.  The resulting FP is practically
indistinguishable from the observed one, as apparent from
Fig. \ref{fig:fp}.  Moreover, also the $L$-$\sigma$ and $L$-$R$
relations of simulated and real clusters are in remarkable agreement
(Fig. \ref{fig:fj}).

Note that assuming a {\it constant} M/L ratio, the FP of simulated
clusters is still well defined, and only shows a mild tilt with
respect to the observed one (see solid line in Fig. \ref{fig:fp}).

\section{Scaling relations at higher redshift for simulated clusters}

While no observational data are yet available for galaxy clusters at
higher redshift, our high-resolution re-simulations allow to
investigate the scaling relations of their DM counterparts at any
epoch.  For that purpose, the most massive progenitors at $z=0.5$ and
at $z=1$ of our 13 DM halos have been considered, and their projected
half-mass radius, velocity dispersion and luminosity have been derived
with the same procedure used for the $z=0$ objects.  Also in this
case, when combining $\rh$, $\sigma$, and $L$ with the coefficients
$\alpha_i$, $\beta_i$, $\gamma_i$ derived for the real clusters at
$z=0$, the resulting FP is very similar to the edge-on FP observed
locally, while when seen face-on, the region populated by the high-$z$
DM halos is systematically shifted towards larger (smaller) values of
$p_1$ ($p_2$) with respect to the region occupied by the observed
nearby clusters.  In addition, very well defined $L$-$\sigma$ and
$L$-$R$ relations are already in place at these redshifts.  With
respect to the observed $L$-$\sigma$ relation at $z=0$, a flatter
slope is found for increasing redshift for the simulated clusters,
while the opposite trend is found in the case of the $L$-$R$ relation,
that becomes steeper for increasing $z$.

%------------------------------------------Two column figure (place early!)
%______________________________________________________________

%------------------------------------------Two column figure (place early!)
%______________________________________________________________

\section{Discussion and conclusions}

We have described a technique for increasing the (mass and spatial)
resolution of objects selected in a given cosmological simulation.
Applying it to sample of 13 massive halos at $z=0$ in a $\Lambda$CDM
cosmology, we have investigated whether the scaling relations observed
for nearby galaxy clusters are also followed by their DM hosts, in a
dissipationless hierarchical merging scenario.  By considering the
most massive progenitors at $z=0.5$ and $z=1$ of the 13 selected
halos, we have also investigated whether the same scaling relations
were already in place at those earlier epochs.

The main conclusions can be summarized as follows:
\begin{itemize}

 \item the high-resolution re-simulation technique is very powerful
and reliable. The precision of the method depends on the number of
particles composing the original object, and on the dimension of the
boundary region considered to avoid the ``contamination'' from
macroparticles.

 \item The DM hosts of galaxy clusters do define a FP that is
practically indistinguishable from the one observed for nearby galaxy
clusters.

 \item A very good agreement between the scaling relations of
simulated and observed clusters is also found in terms of the
$L$-$\sigma$ and the $L$-$R$ relations.

 \item The FP of simulated clusters is already defined at $z=0.5$ and
$z=1$, and it is very similar and as thin as the one observed locally;
the only differences are due to an obvious shift towards smaller radii
and masses for increasing $z$. Also the $L$-$\sigma$ and $L$-$R$
relations are already in place at high redshifts, with a difference in
the slope (with respect to the relations observed locally) due to the
fact that at recent epochs, more massive objects evolve more rapidly
than smaller systems.

\end{itemize}

All the results about the scaling relations of simulated and observed
clusters are obtained by assuming a mass-to-light ratio with mean
value of the order of $450 h \msol/\lsol$ (well in agreement with what
estimated from observations of galaxy clusters), and with a dependence
on the luminosity also suggested by the observations.

Given the very different nature of the two components (baryonic and
non baryonic) and of the physical processes acting on them, the fact
that the DM halos define the same relations shown by real clusters is
not obvious at all.  Moreover, dissipationless hierarchical merging at
galactic scales is unable to reproduce the observed scaling relations
of elliptical galaxies: the end-products of major mergers do lie on
the observed FP, but substantial deviations from the Faber-Jackson and
the Kormendy relations are found in this case (see Londrillo et al.,
this conference).  Such a result suggests that, while dissipation has
a major role in the formation and evolution of ellipticals, it is
negligible with respect to the pure gravity in settling the properties
of galaxy clusters.

\begin{acknowledgements}
B.L. acknowledges G. Tormen, V. Springel, S. White, \& G. Mamon for
their help with the N-body simulations and for useful
discussions. This work has been partially supported by the Italian
Ministery (MIUR) grant COFIN2001 ``Clusters and groups of galaxies:
the interplay between dark and baryonic matter'', and by the Italian
Space Agency grants ASI-I-R-105-00 and ASI-I-I-037-01. L.C. was
supported by COFIN2000.
\end{acknowledgements}

\bibliographystyle{aa}

\end{document}